\documentclass[usegraphicx,twocolumn]{aastex6}

\usepackage{rotating}
\usepackage[3D]{movie15}
\usepackage{hyperref}

\usepackage{url}
\usepackage{times}
\usepackage{mathptmx}
\usepackage{amssymb}
\usepackage[fleqn]{amsmath}
\usepackage{subfigure}
\usepackage{float}
\usepackage{enumitem}
\usepackage{txfonts}
\usepackage{xcolor}
\usepackage{bm}
\usepackage{natbib}
\usepackage{bibentry} 
\usepackage{etoolbox} 
\usepackage{epsfig}
\usepackage{xspace}

\newcommand{\Shape}{\textsc{shape}\xspace}
\newcommand{\Splash}{\textsc{splash}\xspace}

\newcommand{\Blender}{\textsc{blender}\xspace}
\newcommand{\Meshlab}{\textsc{meshlab}\xspace}
\newcommand{\MdotA}{\ifmmode{\dot{M}_{\eta_{\mathrm{A}}}}\else{$\dot{M}_{\eta_{\mathrm{A}}}$}\fi\xspace}
\newcommand{\MdotB}{\ifmmode{\dot{M}_{\eta_{\mathrm{B}}}}\else{$\dot{M}_{\eta_{\mathrm{B}}}$}\fi\xspace}
\newcommand{\etaA}{$\eta_{\mathrm{A}}$\xspace}
\newcommand{\etaB}{$\eta_{\mathrm{B}}$\xspace}
\newcommand{\ec}{$\eta$~Car\xspace}

\makeatletter
\newcommand\bibstyle@comma{\bibpunct{(}{)}{,}{a}{}{,}}
\newcommand\bibstyle@semicolon{\bibpunct{(}{)}{;}{a}{}{,}}
\makeatother

\pretocmd\citet{\citestyle{comma}}\relax\relax
\pretocmd\citep{\citestyle{semicolon}}\relax\relax

\newcommand{\Ms}{\ifmmode{~\mathrm{M}_{\odot}}\else{$\mathrm{M}_{\odot}$}\fi\xspace}
\newcommand{\Msy}{\ifmmode{\Ms \per{yr}}\else {$\Ms \per{yr}$}\fi\xspace}
\newcommand{\Ls}{\ifmmode{~\mathrm{L}_{\odot}}\else{$\mathrm{L}_{\odot}$}\fi\xspace}
\newcommand{\kms}{\ifmmode{~\mathrm{km}\per{s}}\else {$\mathrm{km}\per{s}$}\fi\xspace}
\newcommand{\Rs}{\ifmmode{~\mathrm{R}_{\odot}}\else{$\mathrm{R}_{\odot}$}\fi\xspace}
\newcommand{\per}[1]{\ifmmode{\mathrm{\,#1}^{-1}}\else {$\mathrm{\,#1}^{-1}$}\fi\xspace}

\newcommand{\ten}[1]{\ifmmode{10^{#1}}\else{$10^{#1}$}\fi\xspace}
\newcommand{\sci}[2]{\ifmmode{#1 \times 10^{#2}}\else{$#1 \times 10^{#2}$}\fi\xspace}

\newcommand{\HI}{\ifmmode{\mathrm{H\,I}}\else{H\textsc{$\,$i}}\fi\xspace}
\newcommand{\HII}{\ifmmode{\mathrm{H\,II}}\else{H\textsc{$\,$ii}}\fi\xspace}
\newcommand{\HeI}{\ifmmode{\mathrm{He\,I}}\else{He\textsc{$\,$i}}\fi\xspace}
\newcommand{\HeII}{\ifmmode{\mathrm{He\,II}}\else{He\textsc{$\,$ii}}\fi\xspace}
\newcommand{\HeIII}{\ifmmode{\mathrm{He\,III}}\else{He\textsc{$\,$iii}}\fi\xspace}
\newcommand{\FeIII}{\ifmmode{[\mathrm{Fe\,III}]}\else{[Fe\textsc{$\,$iii}]}\fi\xspace}
\newcommand{\FeII}{\ifmmode{[\mathrm{Fe\,II}]}\else{[Fe\textsc{$\,$ii}]}\fi\xspace}
\newcommand{\NiII}{\ifmmode{[\mathrm{Ni\,II}]}\else{[Ni\textsc{$\,$ii}]}\fi\xspace}
\newcommand{\nuHI}{\ifmmode{\nu_{\HI}}\else{$\nu_{\HI}$}\fi\xspace}
\newcommand{\nuHeI}{\ifmmode{\nu_{\HeI}}\else{$\nu_{\HeI}$}\fi\xspace}
\newcommand{\nuHeII}{\ifmmode{\nu_{\HeII}}\else{$\nu_{\HeII}$}\fi\xspace}

\defcitealias{madura13}{M13}
\defcitealias{clementel14}{C14}
\defcitealias{steffen14}{S14}
\defcitealias{madura15}{M15}

\shorttitle{Astronomical 3-D Printing and \ec}
\shortauthors{Madura}

\begin{document}

\title{A Case Study in Astronomical 3-D Printing: The Mysterious $\eta$~Carinae}

\author{Thomas I. Madura\altaffilmark{1}}
\affil{San Jos\'{e} State University \\
Department of Physics and Astronomy \\
One Washington Square \\
San Jos\'{e}, CA 95192-0106, USA}

\altaffiltext{1}{thomas.madura@sjsu.edu}

\begin{abstract}
3-D printing moves beyond interactive 3-D graphics and provides an excellent tool for both visual and tactile learners, since 3-D printing can now easily communicate complex geometries \emph{and} full color information. Some limitations of interactive 3-D graphics are also alleviated by 3-D printable models, including issues of limited software support, portability, accessibility, and sustainability. We describe the motivations, methods, and results of our work on using 3-D printing (1) to visualize and understand the \ec Homunculus nebula and central binary system and (2) for astronomy outreach and education, specifically, with visually impaired students. One new result we present is the ability to 3-D print full-color models of \ec's colliding stellar winds. We also demonstrate how 3-D printing has helped us communicate our improved understanding of the detailed structure of \ec's Homunculus nebula and central binary colliding stellar winds, and their links to each other. Attached to this article are full-color 3-D printable files of both a red-blue Homunculus model and the \ec colliding stellar winds at orbital phase 1.045. 3-D printing could prove to be vital to how astronomer's reach out and share their work with each other, the public, and new audiences.
\end{abstract}


\section{Introduction} \label{sec:intro}

Three-dimensional (3-D) printing has the potential to provide astronomers with an entirely new means to visualize, understand, and communicate scientific results. 3-D printing or ``additive manufacturing'' has become a useful technique for a wide variety of applications, including prototyping, tissue engineering, materials for energy, chemistry reactionware, molecular visualization, microfluidics, and low-density, high-strength materials \citep{tumbleston15}. In recent years, 3-D printing techniques have advanced beyond the standard processes of fused deposition modeling, selective laser sintering, or stereolithography, and now include full-color sandstone, multi-material digital ABS (Acrylonitrile Butadiene Styrene), and continuous liquid interface production. The decreasing cost and increasing availability of reliable 3-D printers has facilitated access to this technology, and many universities, research institutes, and libraries now have so-called ``Makerspaces'' where nearly anyone can experiment with 3-D printing.

The astronomical community has yet to fully embrace 3-D printing, but the first serious uses of the technology for astronomy and astrophysics research, education, and outreach have already appeared. \citet{christian15} and \citet{grice15} used special software and 3-D printing to transform two-dimensional (2-D) \emph{Hubble Space Telescope} (\emph{HST}) images into tactile 3-D models with the goal of communicating the wonders of astronomy to the blind and visually-impaired. The NASA 3-D Resources web page\footnote{\url{http://nasa3d.arc.nasa.gov/}} offers a growing collection of over 280 free-to-use 3-D models, textures, and images, many of which are 3-D printable. In the professional research literature, \citet{vogt14} 3-D printed a physical model of a 3-D emission line ratio diagram (see their figs.~1 and 15), while \citet{vogt16} 3-D printed a model of the \HI distribution inside HCG~91 (see their fig.~4). \citet[][hereafter S14]{steffen14} included a 3-D printable STereoLithography (STL) file in the online supplemental material for their paper on the 3-D structure of the Eta~Carinae (\ec) Homunculus nebula. Very recently, \citet{clements16} used 3-D printing to produce monochromatic and color representations of the all-sky Cosmic Microwave Background (CMB) intensity anisotropy maps produced by the \emph{Planck} mission.  \citet{clements16} also made the CAD (computer-aided design) files of their models freely available for non-commercial use.

Moving beyond observational data, \citet[][hereafter M15]{madura15} showed that 3-D printing can be used to analyze and communicate results from numerical simulations. \citetalias{madura15} created the first 3-D prints of output from simulations of a multi-dimensional and time-varying astrophysical system; the colliding stellar winds in the massive \ec binary. These 3-D prints helped reveal previously unknown ``finger-like'' structures that occur at orbital phases shortly after periastron, and which protrude radially outward from the spiral wind-wind collision region that forms as the two stars move rapidly about periastron in their highly elliptical (eccentricity $\sim 0.9$) orbits. It is thought that these fingers are related to instabilities that arise at the interface between the dense, radiatively-cooled layer of post-shock primary-star wind and the faster, adiabatic post-shock companion-star wind \citepalias{madura15}. 

Similarly, in late 2015, M. Arag\'{o}n-Calvo of the University of California, Riverside, used 3-D printing to visualize and understand the time evolution of 2-D computer simulations of the cosmic web\footnote{See \url{https://ucrtoday.ucr.edu/33743}}. By taking time to be the third spatial variable, Arag\'{o}n-Calvo was able to more easily identify and track tridimensional cosmic structures in his simulations. \citet{naiman16} has additionally demonstrated how the 3-D modeling software \Blender can be used for astronomy data visualization and the generation of 3-D printable isosurfaces that are based on the results of 3-D simulations (see their fig.~9). More recently, \citet{hendrix16} produced a 3-D print at one time step from their 3-D hydrodynamical simulation of the colliding stellar winds in the Wolf-Rayet 98a (WR~98a) binary system (their fig.~B1).

The 3-D modeling work by \citetalias{steffen14} and \citetalias{madura15} led to the first strong evidence of a direct impact of the central \ec binary, and possibly its colliding stellar winds, on the structure of the surrounding bipolar Homunculus nebula. 3-D printing has helped us to convey this understanding to others. The results of \citetalias{steffen14} and \citetalias{madura15} have also garnered a significant amount of attention from the press and public, and rejuvenated interest in using 3-D printing to help teach astronomy and astrophysics to blind and visually impaired students. However, each of the previously mentioned works illustrates how 3-D printing can be used to help interpret multi-dimensional astrophysical data and share the results with both the broader research community and new audiences. 

Here we describe the motivations, methods, and results of our work on using 3-D printing (1) to visualize and understand the \ec Homunculus nebula and central binary and (2) for astronomy education/outreach, specifically, with the visually impaired. One pivotal result we present is the ability to 3-D print full-color models in which the color, assigned via a pre-defined color table, represents a desired physical quantity (e.g. density). Such full-color models move beyond the simple monochromatic 3-D prints in e.g. \citetalias{madura15} and are generally more informative and useful for visual learners. 

In the following section (\ref{sec:motive}) we summarize the motivations for this work and provide some background on \ec.  We describe our methodology and numerical approach, and the generation of the 3-D printable files in Section~\ref{sec:meth}. Section~\ref{sec:results} presents a brief description of the results including standard 2-D images, pictures, and interactive 3-D graphics. A discussion of the 3-D printing results and their use for education and outreach, including with the visually impaired, is in Section~\ref{sec:disc}. Section~\ref{sec:summ} summarizes our conclusions.

\section{Motivation \& Background} \label{sec:motive}

\subsection{The Need for Multi-D Visualization and Tactile Models} \label{ssec:need}

The Universe is at least four-dimensional (4-D), yet historically, most astronomical data have consisted of 2-D images or 1-D, sometimes 2-D, spectra/time series. Unfortunately, we cannot send probes to study astrophysical (extrasolar) objects from the multiple viewing angles needed to provide us with full 3-D spatial information. Instead, much of what we know about the Universe comes from our understanding of the complex interactions of matter with radiation.
 
In recent decades, a combination of spectral and spatial mapping has allowed astronomers to determine the 3-D spatial nature of cosmic systems, and sometimes their changes with time, if velocities can be unambiguously converted to distances. Spatially-resolved spectra can be obtained by stepping a spectrographic slit across a target and recording exposures at each position, producing a 3-D (2-D spatial + 1-D wavelength or velocity) data cube . Data cubes at different epochs provide 4-D information. The underlying 3-D structure and properties of a system can be teased out by comparing the observational data cubes to synthetic cubes generated using 3-D simulations, which predict extended emitting structures as a function of underlying physical parameters.

Such spatial and spectral mapping techniques can be very time consuming, since stepping a spectrographic slit across a target or field requires moving the telescope to each necessary position. However, the rapidly growing field of integral field spectroscopy (IFS) allows one to image the entirety of an astronomical object/field and obtain a full spectrum at each pixel in the image simultaneously. IFS is now used at all major Earth observatories, and soon the near-infrared spectrograph on the James Webb Space Telescope (JWST) will be the first multi-object spectrograph in space. The number of data cubes from IFSs is likely to increase in coming years. Discovering innovative ways to visualize, understand, and communicate the contents of such multi-D datasets is essential if we are to make sense of the intrinsically multi-D Universe in which we live.

The predominance of 2-D figures and animations in the astrophysical literature is likely historically driven by the fact that astronomical images are recorded on a 2-D surface (photographic plate or CCD detector). Early numerical simulations were also limited to 1, 2, or 2.5 dimensions until computers became powerful enough to permit realistic physical modeling in 3-D on a reasonable time scale. While astronomical data sets and numerical simulations have evolved beyond two dimensions, only relatively recently have astronomical journals started supporting 3-D figures. Prior to this, authors were limited to the classic 2-D paper journal format. However, the reduction of intrinsically multi-D data to two dimensions, or the projection of a 3-D image to 2-D, typically implies a loss of information. Such information loss is not always harmful, as it can help highlight more important aspects of a dataset. Nonetheless, there is a growing need for researchers to move beyond the inherent limits of 2-D displays, both for their own understanding and for communicating with colleagues, funding agencies, and the public.

There is no compelling reason why today's researchers should be limited to 2-D graphics when interpreting or sharing their results. This is especially true since all major astrophysical journals are currently published online. ``Augmented articles'' \citep{vogt13} are now possible, in which interactive 3-D models, images, sounds, and videos are included directly within the \emph{Adobe Portable Document Format} (PDF) article. The use of interactive 3-D models in the astrophysics literature, via methods such as those described in \citet{barnes08}, in which the \textsc{S2PLOT} programming library and Adobe's 3-D extensions to PDF are used to create and incorporate interactive 3-D models into PDF documents, are slowly becoming popular, and multiple astrophysical journals fully support the inclusion of interactive 3-D figures. Some recent examples of such interactive 3-D PDF figures can be found in \citet{vogt14}, \citetalias{steffen14}, and \citetalias{madura15}. More recently, \citet{vogt16} outlined the concept of the ``X3D pathway'' as a means of simplifying and easing the access to data visualization and publication via interactive 3-D diagrams.

While interactive 3-D graphics will likely prove to be extremely useful to astronomers and the public, the exploration of alternative and emerging technologies for data visualization and communication is always warranted. This is especially true as astronomers attempt to reach out and share their work with new audiences, some of which have been historically neglected by the astronomical community. One such community is the visually impaired, who are seriously underrepresented in science, but have an avid interest in astronomy \citep[][and references therein]{christian15}. Creating learning materials accessible to a wide audience with different learning styles and challenges is often difficult, time consuming, and expensive, which is why the astronomical material currently accessible to those who are visually impaired is very limited \citep{grice15}. 3-D printing offers an innovative means to reach out to tactile learners, particularly the visually impaired, and provides a unique way to translate astronomical data and numerical simulations into tactile models that can assist individuals in building mental images of the Universe around them \citep{grice15}.

\subsection{Background on \ec} \label{ssec:etacar}
\vspace{2.0mm}
\subsubsection{The Homunculus Nebula} \label{sssec:homunc}

In the mid 1840s, \ec produced the greatest non-terminal stellar explosion ever recorded \citep{davidson97}, during which it became the second brightest non-Solar-System object in the sky and ejected 10 -- 40 \Ms of material, forming the bipolar nebula known as the ``Homunculus'' \citep{smith03,gomez10}. Surprisingly, this did not destroy the central star(s). Observations reveal that \ec itself is currently a massive ($\gtrsim$120~\Ms \citealt{hillier01,hillier06}) evolved binary with a highly eccentric ($e \sim 0.9$), 5.54-year orbit \citep{damineli08a,damineli08b}. At a distance of 2.3~kpc \citep{smith06}, \ec is our closest and most luminous ($L_{\mathrm{Total}} \gtrsim \sci{5}{6} \Ls$, \citealt{davidson97, hillier01, hillier06}) example of an evolved supermassive star and supernova progenitor that can be studied in great detail.

Numerous explanations have been proposed for \ec's Great Eruption and the formation of the Homunculus (see many references in e.g. \citetalias{steffen14}). If the central binary influenced the formation or shaping of the Homunculus, it should have left an imprint in the Homunculus. There are at least two ways binarity and the Homunculus's formation might be related. First, binarity could be directly responsible for triggering the eruption, via a stellar collision or merger \citep{smith11,pz16}. Second, the binary could influence the shape of the nebula during and/or after any explosion, perhaps via the companion star's extremely fast, low-density wind and orbital motion. 

The apparent 3-D alignment of the orbital axis (orthogonal to the orbital plane and through the system center of mass) of the \ec binary with the inferred polar symmetry axis of the Homunculus appears to be strong evidence for some sort of binary influence on the formation/shaping of the Homunculus \citep{madura12}.  However, direct evidence of binary interaction in the Homunculus would shed more light on the eruption process and shaping mechanism, as well as help to further constrain the orbital orientation. In order to find such evidence, the structure of the Homunculus needed to be described in more detail than that provided by simple axisymmetric models. The detailed 3-D modeling of the Homunculus presented in \citetalias{steffen14}, based on ESO Very Large Telescope (VLT)/X-Shooter spectral mapping observations of the H$_{2}$ $\lambda = 2.12125$~$\mu$m emission line, was the first to achieve this, and the first to reveal important deviations from the axisymmetric bipolar morphology. 3-D printing the 3-D Homunculus model by \citetalias{steffen14} provides us with a unique way to visualize the nebula, interpret its features, and share the 3-D model with colleagues, the public, and the visually impaired. 

\newpage
\subsubsection{The Massive Colliding Wind Binary} \label{sssec:binary}

Due to their extreme luminosities, the stars in \ec have powerful radiation-driven stellar wind mass outflows. The Luminous Blue Variable (LBV) primary component, \etaA, has perhaps the densest known stellar wind ($\MdotA \approx \sci{8.5}{-4} \Msy$, $v_{\infty} \approx 420 \kms$; \citealt{hillier01, groh12a}), while the less luminous companion star, \etaB, has a much lower density, but incredibly fast wind ($\MdotB \approx \sci{1.4}{-5} \Msy$, $v_{\infty} \approx 3000 \kms$; \citealt{pittard02, parkin09}). These winds collide violently, producing a series of shocks and wind-wind interaction regions (WWIRs) that are the sources of numerous forms of time-variable emission and absorption seen across a wide range of wavelengths \citep{damineli08b}. Due to the high orbital eccentricity and hot ($\sim$40,000~K) \etaB, the WWIRs in \ec produce dense spiral structures of compressed, irradiated gas that are spatially and spectrally resolved by the \emph{Hubble Space Telescope}/Space Telescope Imaging Spectrograph (\emph{HST}/STIS) in numerous forbidden emission lines \citep{gull09,gull11,teodoro13,gull16}.

Properly modeling the \ec binary and its WWIRs requires a 3-D time-dependent treatment since orbital motion affects the geometry and dynamics of the WWIRs.  3-D hydrodynamical simulations have helped to greatly increase our understanding of the \ec binary and its WWIRs, and how they affect numerous observational diagnostics of the system (see e.g. \citealt{okazaki08}, \citealt{parkin11}, \citealt{maduragroh12}, \citealt{madura12, madura13}, \citealt{richardson16}). The ability to adequately visualize the full 3-D time-dependent geometry of such simulations is a crucial component to studies of the \ec system, especially if we are to better understand how and where various forms of observed emission and absorption originate, and thus the history and evolution of this nearby massive supernova progenitor.

Most published figures of 3-D simulations of \ec's binary colliding winds, and colliding wind binaries in general, consist of 2-D slices through the 3-D simulation domain, with color representing a scalar quantity such as density or temperature (see \citealt{okazaki08}, \citealt{parkin11}, and \citealt{madura13}, hereafter M13). While multi-panel figures showing 2-D slices can be useful, the time-varying geometry of the WWIRs combined with parameter studies of various stellar, wind, and orbital parameters, can lead to large numbers of cumbersome 2-D figures. Furthermore, the amount of information that 2-D slices can convey about an intrinsically 3-D structure is limited, which can make the interpretation of 2-D slices difficult.

As with the \citetalias{steffen14} Homunculus model, 3-D printing provides a new method for visualizing and studying \ec's binary system and WWIRs. 3-D printing moves beyond 2-D slices and interactive 3-D graphics and provides an excellent tool for visual \emph{and} tactile learners. 3-D printable models also help alleviate some of the current limitations of interactive 3-D figures, such as limited software support and difficulty of file generation (many free programs exist for generating 3-D print files under a common standard), portability (3-D print files typically share the same universal file formats, have small file sizes, and can be easily hosted in online repositories), accessibility (free software can be used to view or modify the files, plus 3-D printed models are accessible to the visually impaired), and potential long term sustainability (the most common file format has been around for decades and will likely continue to be supported as 3-D printing grows in popularity and affordability).

\section{Methods} \label{sec:meth}

\subsection{3-D Printing the \ec Homunculus}\label{ssec:homunc3Dprint}
\vspace{2.0mm}
\subsubsection{Monochromatic 3-D Prints}\label{sssec:homuncmono}

In 2012 March, the entire Homunculus nebula (14~arcsec $\times$ 20~arcsec) was mapped using the ESO VLT/X-shooter spectrograph (details of the observations and data reduction are in \citetalias{steffen14}).  By modeling a specific single emission line of molecular hydrogen (H$_{2}$) at 2.12125~$\mu$m using the virtual astrophysical modeling laboratory \Shape (version 5, \citealt{steffen11}), \citetalias{steffen14} was able to construct the first full 3-D model of the Homunculus nebula that includes small-scale structures. Additional specific details on the 3-D modeling, results, and conclusions can be found in \citetalias{steffen14}, but this improved 3-D model of the Homunculus revealed, for the first time, important deviations from the axisymmetric bipolar morphology, which suggest that the central massive binary played a significant role in shaping the Homunculus.  For improved visualization, a 3-D mesh structure for the Homunculus was generated and an interactive version of the 3-D model was included in fig.~5 of the \citetalias{steffen14} PDF publication.

\begin{figure*}
 \begin{center}
    \includegraphics[width=177mm]{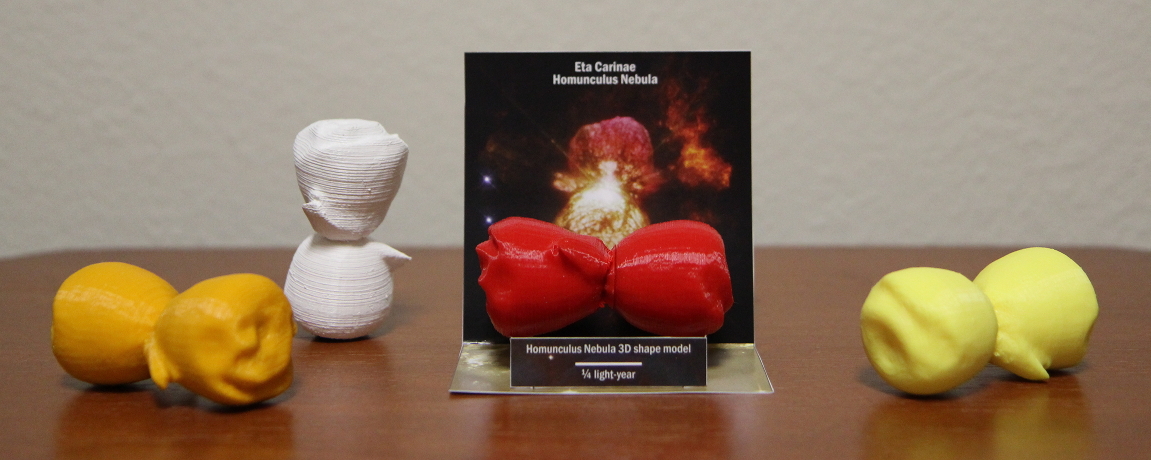}
    \caption{Example monochromatic 3-D prints of \ec's Homunculus nebula, based on the 3-D model by \citetalias{steffen14}$^{5}$.}\label{fig1}
 \end{center}
\end{figure*}

Once a 3-D mesh structure for the Homunculus was generated, it could be written to the standard STL file format for 3-D printing (the original STL file is included as supplementary material to \citetalias{steffen14}). This file can be imported into a software program such as \Blender\footnote{\url{http://www.blender.org}} or \Meshlab\footnote{\url{http://meshlab.sourceforge.net/}} to detect and correct non-manifoldness, mesh holes, loose objects, and inverted normals, as is required for a successful 3-D print. It is important that any 3-D design file to be 3-D printed be closed or ``watertight," and that all components in the 3-D model be connected to create a solid and have a finite thickness that is capable of being printed by the desired 3-D printer (i.e. features in the model must be of the appropriate size for the resolution of the printer and not too small). The model furthermore needs to be manifold (have no edges that are shared between more than two faces) and any individual/floating points, lines, and planes should be removed. Surface normals must all be pointing in the same direction outward from the surface of the model to ensure that the 3-D printer does not confuse the internal and outer surfaces. Finally, before printing, the 3-D model should be scaled to the appropriate dimensions so that it will fit on the 3-D printer bed.

The watertight and manifold mesh requirements may limit the sizes and types of structures that can be 3-D printed, so care must be taken during the mesh creation phase to ensure a printable model. Non-manifold edges are parts of the model where the mesh points have not been connected to create a solid. Isolated or disconnected points, lines, and planes do not have the three separate dimensions necessary for 3-D printing and will result in either a slicing-software error or a failed print. This is a common occurrence during the design phase when two components appear ``connected'' on the screen, but in reality, the seams between the two objects have not been joined. In some cases, components of the model may overlap one another, confusing the slicing engine and leading to an unprintable model. Extremely high resolution or small details in a model may also sometimes not be printable, or may be printed inaccurately, because of the manifold and watertight requirements.

As a result, it becomes increasingly difficult to 3-D print any model that contains non-3D structures or a lot of ``free-floating'' structures that are not attached to each other or to anything else. Free-floating components require supports, which will likely need to be added during the mesh creation phase (see e.g. fig.~4 of \citealt{vogt16}).

While different 3-D printers and 3-D ``slicing'' software may have more stringent requirements, we found that generally, the above were the minimal requirements necessary for a suitable STL file and successful 3-D print. In the consumer market, most 3-D printers are monochromatic and contain a single extruder. As such, the color of the 3-D model might not matter, especially when one is only concerned with the overall 3-D geometry of the object, as was initially the case for the Homunculus. A colorless model is also appropriate when the target audience consists of blind or visually impaired learners. In such cases, an STL file is sufficient, especially since color information is usually not stored in the STL file format. The choice of color is then a manual decision, and the user simply picks their favorite color of filament to feed into their 3-D printer. Fig.~5 of \citetalias{steffen14} presents the original Homunculus nebula model, while Fig.~\ref{fig1} shows example monochromatic 3-D prints of the Homunculus\footnote{The STL file of this model is available at \url{https://nasa3d.arc.nasa.gov/detail/eta-carinae-homunculus-nebula}}.

\subsubsection{Multi-Color 3-D Prints}\label{sssec:homunccolor}

Although 3-D printers have historically been single material and single color, recent advancements allow for mutli-color and multi-material 3-D printing. This technology is not limited to large, expensive professional 3-D printers either, as a number of consumer-grade 3-D printers are available that contain multiple extruders and allow for simultaneous mutli-color printing. The simplest example would be a desktop 3-D printer with two extruders that permit printing in two colors simultaneously. At the more extreme (expensive) end are professional full-color 3-D printers, full-color sandstone being the most widely available example at this time.

The ability to 3-D print in different materials and colors opens up a plethora of design choices for researchers and educators, even if only geometry information is initially available. As an example, let us explore one method of augmenting the 3-D Homunculus model by simply adding some color. 

\begin{figure}
\centering \includemovie[
        toolbar, 
        label=Fig2.u3d,
     text={\includegraphics[width=82mm]{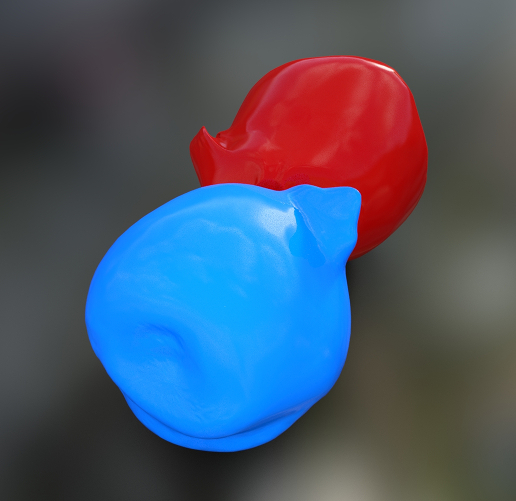}},
     3Daac=60.000000, 3Droll=-5.500000, 3Dc2c=-0.7478541135787964 -0.27854451537132263 0.6026003956794739, 3Droo=420.000000, 3Dcoo=-0.5884353518486023 5.588916778564453 -43.42313003540039,
        3Dlights=CAD,
]{}{}{Fig2.u3d}
\caption{Red-blue \ec Homunculus nebula model, based on the 3-D model of \citetalias{steffen14}. The blue lobe is the approaching foreground lobe, while the red lobe is the receding background lobe, as seen in e.g. \emph{HST} images of \ec. Click the image for an interactive 3-D version (Adobe Reader$^{\circledR}$ only).}
\label{fig2}
\end{figure}

One shortcoming of the monochromatic 3-D Homunculus model is that it is difficult to distinguish which of the bipolar lobes is the approaching foreground lobe (as seen in images) and which is the receding background lobe. However, using free 3-D creation software such as \Blender, it is straightforward to add color to each lobe in the Homunculus model so as to more easily identify which lobe is which. The choice of color can also be based on physical properties of each lobe, providing additional information to the user. For instance, we can color the approaching foreground lobe blue, since it is the ``blue-shifted'' lobe, and the receding background lobe red, since it is the ``red-shifted lobe.''  Each colored lobe can then be saved as an STL file for use with a simple dual-extruder, dual-color 3-D printer, or the complete full-color model can be exported in the VRML, WRL, or X3D file format, which can then be 3-D printed in full-color sandstone. Fig.~\ref{fig2} presents such a red-blue color-coded 3-D printable Homunculus model, created with \Blender.\\

\subsection{3-D Printing \ec's Binary Colliding Stellar Winds}\label{ssec:binary3D}
\vspace{2.0mm}
\subsubsection{The 3-D SPH Simulations}\label{ssec:SPH}

The 3-D hydrodynamical simulation snapshots 3-D printed for \citetalias{madura15} correspond to specific phases (apastron, periastron, and 3~months after periastron) from the Case~A ($\MdotA \approx \sci{8.5}{-4} \Msy$) and Case~C ($\MdotA \approx \sci{2.4}{-4} \Msy$) small-domain ($r = 10\,a = 155$~AU) 3-D SPH simulations of \citetalias{madura13}. Here we discuss only the Case~A simulation results as they are more likely representative of the primary star's current mass loss rate \citep{madura13, clementel15a, clementel15b, teodoro16, gull16}. The primary star's wind terminal speed is set to $420 \kms$, while the companion star mass loss rate and wind terminal speed are set to $\MdotB = \sci{1.4}{-5} \Msy$ and $v_{\infty} = 3000 \kms$, respectively. The SPH particle mass is $5.913 \times 10^{24}$~g for the wind of \etaA and $5.913 \times 10^{23}$~g for the wind of \etaB in Case~A. We refer the reader to \citetalias{madura13} and references therein for further details on the SPH simulations and results. 

The spherical computational domain radius $r = 155$~AU was chosen in order to investigate at high spatial resolution the structure of \ec's inner WWIRs, since the ``current'' interaction between the two winds occurs at scales comparable to the semi-major axis length $a \approx 15.4$~AU. The three orbital phases selected represent when the WWIR has its simplest (apastron) and most complex (periastron and 3~months after periastron) geometries. The snapshot at 3~months after periastron ($\phi = 1.045$) corresponds to when the WWIR has a distinct Archimedean-spiral-like shape in the orbital plane due to the rapid orbital motion of the stars around periastron \citep{okazaki08, parkin11, madura12, madura13}. A standard $xyz$ Cartesian coordinate system is used, with the orbit set in the $xy$ plane, the origin at the system center of mass, and the major axis along the $x$-axis. The stars orbit counter-clockwise when viewed from above along the $+z$-axis. Periastron is defined as $t = 0$ ($\phi = t/2024 = 0$) and simulations are started at apastron.

\subsubsection{Grid Construction and Density Distribution}\label{ssec:Grid}

As SPH is a mesh-free method for solving the equations of fluid dynamics \citep{monaghan05}, we must find some way to generate an appropriate 3-D mesh in order to 3-D print the SPH simulation output. The difficulty, however, is that the SPH data are highly adaptive and unstructured, defined on a set of points that follow the fluid motion. We chose to generate an unstructured tetrahedral mesh from the SPH particle data using the same methodology as \citet{clementel14, clementel15a}. Using the SPH particles as the generating nuclei, we tessellate space according to the Voronoi recipe, wherein all points in a grid cell are closer to the nucleus of that cell than to any other nucleus. The Voronoi nuclei are then connected by a Delaunay triangulation.

To the nucleus of each Voronoi cell we assign the corresponding SPH particle mass, density, temperature, and velocity, computed using the standard SPH cubic spline kernel, which helps ensure that each scalar variable visualized on our mesh closely matches that of the original SPH simulations. Comparing to a direct visualization of the SPH density using \Splash \citep{price07} shows that this approach indeed matches well the density distribution of the original SPH simulations (see e.g. fig.~1 of \citealt{clementel15a}).

\subsubsection{Visualization}\label{ssec:Vis}

\begin{figure*}
\centering \includemovie[
     3Dviews2=views_color.tex,
        toolbar, 
        label=Fig3.u3d,
     text={\includegraphics[width=177mm]{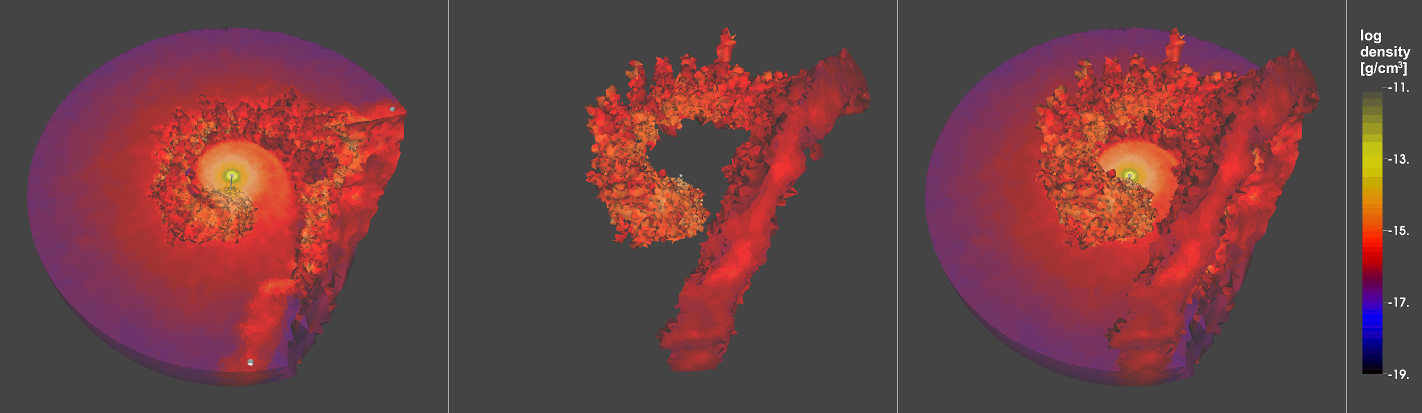}},
        3Dlights=CAD,
]{}{}{Fig3.u3d}
\caption{Full-color visualization of \ec's binary colliding stellar winds and WWIR at orbital phase 1.045 (3 months after periastron), based on the results of the 3-D SPH simulations of \citetalias{madura13}. Color represents logarithm of the density in cgs units. The left panel shows the bottom half of the model only, with the stars, dense \etaA wind, and hollow \etaB wind cavity. The middle panel shows only the WWIR and the stars. The right panel shows the two pieces together, with the WWIR on top.  All views are looking down on the orbital plane. Click the figure for an interactive 3-D version of the model (Adobe Reader$^{\circledR}$ only). Pre-programmed views, available under the ``Views" menu in the 3-D model toolbar, include the complete joined model (view ``Joined"); the \etaA winds, \etaB wind cavity, and stars only (view ``PrimaryWind"); and the WWIR plus stars only (view ``WWIR"). In this and the following figures, the stars orbit counter-clockwise when looking down on the orbital plane. While in interactive mode, right-click and select ``disable content'' to return to the 2-D version of the figure.}
\label{fig3}
\end{figure*}

Since the Delaunay cells of our 3-D mesh are tetrahedra, the quantity we visualize is the average of the four vertices that define the tetrahedron cell (i.e. the average of the four Voronoi nuclei). This works well for visualizing most physical quantities (e.g. temperature, density, velocity). Unfortunately, the fully rendered wind of \etaB prevents one from seeing the complete 3-D geometry of the cavity carved within \etaA's wind by \etaB's wind. It would be better if we could visualize the modified wind of \etaA separate from the lower-density \etaB wind. Accomplishing this turns out to be quite straightforward since \etaA and \etaB have very different values of $\dot{M}$ (\MdotA/\MdotB$\approx 60$), and the SPH simulations use different particle masses for each stellar wind to account for this $\dot{M}$ difference. Using the SPH particle mass we thus isolate the pre- and post-shock \etaA winds while keeping the entire pre- and post-shock \etaB winds, and the top ($z>0$) half of the model, transparent (which reveals the orbital plane). An example of this view, with color showing the logarithm of the density, is shown in the left panel of Fig.~\ref{fig3}.

In order to study the 3-D structure of the spiral WWIR, we must find a way to separate it from the individual stellar winds. As discussed in \citetalias{madura15}, we chose to use temperature and density to isolate the thin, dense post-shock \etaA wind region located between the contact discontinuity and the pre-shock \etaA wind. Radiative cooling of the post-shock \etaA gas increases its density by at least an order of magnitude (\citealt{parkin11}; \citetalias{madura13}), creating a large density contrast between the pre- and post-shock \etaA winds. Because the SPH simulations use constant, spherical mass loss rates, the density dependence with radius from the stellar surface in each pre-shock stellar wind is roughly given by $\rho_{\mathrm{spherical}}(r) = \dot{M} / [4 \pi r^{2} v(r)]$, where $v(r) = v_{\infty}(1 - R_{\star} / r)^{\beta}$ is the standard `beta-velocity law' ($\beta = 1$ for our simulations), with $v_{\infty}$ the wind terminal velocity and $R_{\star}$ the stellar radius \citepalias{madura13,madura15}. Defining $\delta \equiv \rho_{\mathrm{SPH}} / \rho_{\mathrm{spherical}}$, we calculate at each location the contrast in density between what is provided in the SPH simulations ($\rho_{\mathrm{SPH}}$), and what the density at that location should be in an undisturbed spherical stellar wind. We isolate the post-shock \etaA wind (from here on referred to simply as the WWIR) by computing $\delta$ and selecting only those regions with $\delta > 2$ and $T = 10,000$~K, since as the post-shock \etaA wind cools radiatively, it remains at the floor temperature $T = 10,000$~K set in the SPH simulations \citepalias{madura13}. Considering only regions with $T = 10,000$~K also helps to isolate the post-shock \etaA wind from the much hotter ($T > \ten{6}$~K) post-shock \etaB gas. The middle panel of Fig.~\ref{fig3} illustrates a view of the 3-D spiral WWIR surface that exists above the orbital plane at orbital phase 1.045.

\subsubsection{Generation of 3-D Print Files}\label{ssec:3Dprintfiles}
 {\centering \paragraph{Monochromatic 3-D Prints} ~\\}

We begin the file generation process by visualizing our unstructured grid data using an appropriate program such as \textsc{VisIt}\footnote{\url{https://wci.llnl.gov/simulation/computer-codes/visit/}} or \text{ParaView}\footnote{\url{http://www.paraview.org/}}. We then use the above-described procedures to isolate either the winds of \etaA or the WWIR. Since we are using a scientific visualization program, we are free to choose any method and color table to display our physical quantity of interest (e.g. logarithm of the density). Next, in order to ensure that our models will fit within the dimensions of the available 3-D printers, while at the same time preserving as much detail as possible, we crop the outer spherical edge of the simulation and choose to visualize and print only the region extending to radius $r = 7a \approx 108$~AU from the binary system center-of-mass. This cropping removes the outer $\sim30$\% of the simulation, but for the orbital phases of interest, no crucial information is lost since this portion of the simulation simply shows the outward propagation of any internal structures formed at earlier phases (see e.g. figs. in \citetalias{madura13}).
\newpage
Once the visualization is created, it is exported as an X3D file, which has the benefit of preserving not only the detailed geometry of the 3-D model mesh, but also the full color information for the selected color table. The X3D file is then imported into \Blender, which has useful tools for detecting and correcting non-manifoldness, mesh holes, loose objects, and inverted normals, which is necessary for the creation of a 3-D printable file. \Blender also allows textures, colors, and meshes to be improved, added, subtracted, etc. 

We made one mesh adjustment in \Blender before exporting the models to the STL format. This modification helps to ensure that our models fit within the 3-D printer and have a stable base on which to stand once printed. To provide an interesting scientific reference point for the 3-D printed models, we rotated them to the correct orientation that the \ec binary has on the sky as seen from Earth, with inclination $i = 138^{\circ}$, argument of periapsis $\omega = 263^{\circ}$, and position angle on the sky of the orbital angular momentum axis of PA$_{z} = 317^{\circ}$ (\citealt{madura12}; \citetalias{madura13}). Once rotated, we then removed a small portion of the bottom of the models so that they have a flat base. We were careful not to remove any of the WWIR or cavity carved within \etaA's wind by \etaB; only a small portion of the undisturbed spherical \etaA wind was removed, just enough to give the models a flat base. This has no effect on the results, but allows the 3-D printed models to be placed on a flat surface and oriented to provide the viewer with an idea of how the binary and WWIR appear on the sky, assuming North is up and East is left.

{\centering \paragraph{Full-Color 3-D Prints} ~\\}

For the initial monochromatic 3-D prints, our main concern was preserving the overall 3-D geometry of the design to be printed. Color was simply an aesthetic choice, especially since a full-color 3-D printer was unavailable at the time, and we were interested in whether a much lower-cost consumer-grade 3-D printer was capable of printing the models with sufficient speed and accuracy. However, 3-D printing in full-color sandstone (FCS) is now commonplace and possible via a large selection of online 3-D printing companies.

FCS printers work layer by layer, spreading on each pass a thin layer of gypsum-like powder and selectively binding it to create the 3-D model. The unique feature of FCS is that it uses CMYK (cyan, magenta, yellow, key/black) printheads and simultaneously binds and dyes to produce the full-color result. Since the ink is absorbed by the powder, the effective resolution of the colors in an FCS model is about 50 DPI, even though the printhead's resolution is about 300 DPI. This is really important to take into consideration when designing very small parts with fine details.

The model's wall thickness also plays a crucial role in determining the colors of an FCS print. FCS printers tend to only dye the outer 1 mm layer of the model, and after the completion of the printing process, the model gets coated with a cyanoacrylate glue that is absorbed by the model and brightens the colors. Thus, if the wall thickness is not consistent, the 3-D model will absorb different amounts of glue on each section, resulting in slight tone differences. Finally, because the colors used in FCS are similar to those in inkjet printers, colors can fade significantly if the model gets wet.

A crucial benefit of the FCS process is that, because FCS is printed by binding together powdered material using a binding agent, the powder acts as support material while building up the print. Therefore, no support structures are needed, providing additional form freedom. This is critical for a model part such as \ec's WWIR, which contains many free-hanging unsupported details that require dissolvable support material when using a standard FDM desktop 3-D printer (such as the WWIR fingers seen at phases near 1.045). The FCS process is also much faster, in our case by about an order of magnitude.

Creating a full-color 3-D print file is straightforward and follows the same methodology described above for generating a monochromatic 3-D print file. The key exception is that after an X3D file is imported into \Blender and the mesh is corrected and additional components are added, instead of exporting the model to the standard STL format, the model is exported as either an X3D, WRL, or VRML file (depending on the model of printer), which again preserves both the mesh geometry and the color. The color information is transferred to the 3-D mesh using the ``Vertex Color Paint'' option under \Blender's ``Material'' button in the ``Properties'' editor. The colors are then made visible in \Blender by ensuring that the ``Textured Solid'' box is checked under the ``Shading'' panel in the ``Properties Region''.

Unlike the monochromatic prints, we did not remove a small portion of the bottom of the models to act as a stand. Rather, we kept the entire outer spherical edge of the primary winds. One modification we did make to the full-color model was the addition of a pair of small pegs to the underside of the WWIR component and a pair of matching aligned holes in the top of the \etaA winds component, so that the two components would fit snugly together (but still be separated if desired) after printing (see Figs.~\ref{fig3} and \ref{fig8}). Also, because the FCS process can easily print unsupported edges/components, instead of having two separate files (one for each component of the model), we placed both model components together in one file, with one piece vertically above the other and a small gap in between. This allowed both components to be printed simultaneously, which resulted in a shorter total print time and ensured that both components were scaled to the same dimensions so that they fit together near perfectly after printing. For our full-color test, we chose to 3-D print only the most dynamically interesting model at orbital phase 1.045, due to the high cost of printing such a large model with FCS ($\sim20$ cubic inches of material).
\newpage
\subsubsection{3-D Printing the Colliding Winds and WWIR}

To print the 3-D monochromatic models, we used a consumer-grade MakerBot Replicator 2X 3-D Printer, which has dual-extrusion. Since the 3-D models are incredibly complex and contain free-hanging unsupported edges, we printed the models in one color using one extruder and used the second extruder to print dissolvable support material. Once printed, we placed the models in a limonene bath that safely dissolves away the support material. We used the highest layer resolution possible (100~microns) and a physical size that occupies nearly the entire available build volume (model diameter $\approx 6$~inches $\approx 15$~cm across its widest part).

The monochromatic 3-D printed models consist of two parts joined by small metal pins that are inserted after printing. The bottom half of each model represents the dense \etaA wind and hollow \etaB wind cavity, while the top half is solely of the WWIR (dense post-shock \etaA wind region). The two pieces are separable and allow the viewer to see the internal details of the cavity carved by \etaB in \etaA's wind, or the WWIR by itself (see e.g. Figs.~\ref{fig5} -- \ref{fig7}). 

Additionally, to help guide the user, we added two small beads connected by a short pin to represent the individual stars and to illustrate their orientation and separation with respect to each other, the dense \etaA winds, and the WWIR. The locations and separation of the beads are to scale; however, in order to help make the individual stars more visible at the scale of our models, we increased the radius of each star to be a factor 3.5 times larger than the correctly-scaled stellar radius. Hence, the beads depicting the stars have radii of 210~\Rs and 105~\Rs for \etaA and \etaB, respectively.

The FCS print was performed courtesy of WhiteClouds, Inc., in April 2015, on a 3DSystems ProJet 600Pro FCS 3-D printer\footnote{See \url{https://www.whiteclouds.com/press-releases//nasa-teams-with-whiteclouds-to-3d-print-eta-carinae-model} and \url{https://3dprint.com/59097/star-model-whiteclouds-nasa/}}. The ProJet 600Pro incorporates professional 4-channel CMYK full-color 3-D printing to a resolution of 600 $\times$ 540 DPI with a minimum feature size of 0.004 inches. For our print, a gamma color table showing the logarithm of the density in cgs units was used. The total print time was between four and five hours, plus approximately two hours of post-processing (cleanup). The model diameter was again $\approx 6$~inches. After printing, a small bead representing \etaB (scaled radius 105~\Rs) was added, with a small pin of correctly-scaled length used to mark its separation from \etaA and the WWIR.

\section{Results} \label{sec:results}

\subsection{The Homunculus}\label{ssec:homuncres}
\vspace{2.0mm}

\begin{figure*}
 \begin{center}
    \includegraphics[width=177mm]{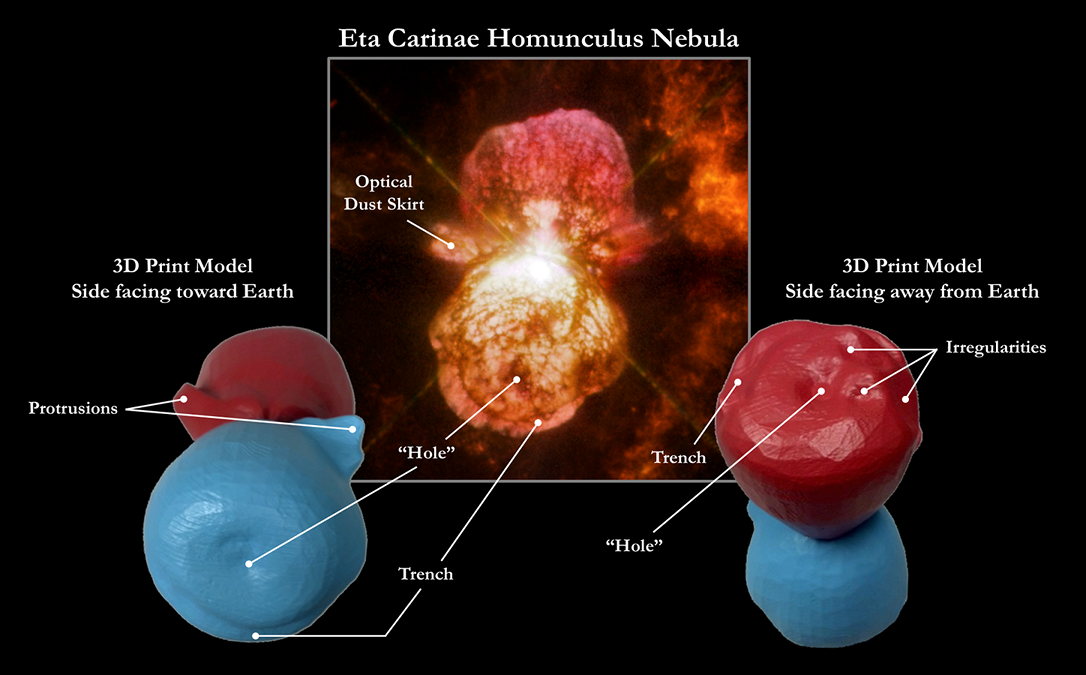}
    \caption{Example dual-color \ec Homunculus nebula 3-D print, based on the 3-D model by \citetalias{steffen14}. Inset: Optical \emph{HST} image of the Homunculus (Inset Credit: NASA, ESA, Hubble SM4 ERO Team).}\label{fig4}
 \end{center}
\end{figure*}

The 3-D prints of the Homunculus help highlight the new small details within the bipolar morphology of the nebula that were identified by \citetalias{steffen14} in their 3-D modeling. Fig.~\ref{fig4} labels in the color version of the 3-D print each of the major features found by \citetalias{steffen14} that show significant deviations from the nebula's overall bipolar shape. For reference, an optical \emph{HST} image of the Homunculus is included. Each of the features is easily found in monochromatic 3-D prints of the Homunculus (see Fig.~\ref{fig1}), but the dual-color version helps to better guide the following discussion.

Near the center of the bottom of each lobe are the indentations or ``holes'' also previously identified by e.g. \citet{teodoro08} in maps of the \FeII 1.257$~\mu$m emission line. The holes are noticeably off-center from the bipolar symmetry axis (by $\sim$8$^{\circ}$, \citetalias{steffen14}). Each polar cap also contains a well-defined trench positioned with approximate point-symmetry with respect to the central stars. Each trench is not quite a half-circle, encompassing a total angle of $\sim$120$^{\circ}$ -- 140$^{\circ}$ \citepalias{steffen14}. The blue trench is seen in optical images of the nebula as a curved dark line that crosses the polar region of the approaching lobe.

There are also two protrusions, one on each lobe, near the mid-plane of the nebula on the Earth-facing side. One protrusion is located to the right (west on the sky) on the blue lobe, while the other is located to the left (north on the sky in images due to the nebula's tilt from the plane of the sky by $\sim$48$^{\circ}$ and its position angle of $\sim$312$^{\circ}$) on the receding red lobe. While the detailed structure of each protrusion is different, both straddle the direction of binary apastron by angles of $\sim$55$^{\circ}$ on either side, as projected onto the orbital plane \citepalias{steffen14}. There are no protrusions on the side of the nebula that faces away from Earth, the side which corresponds to the direction of binary periastron.  The modeling of \citetalias{steffen14} shows that the protrusions are not part of any equatorial outflow; they are outside of the equatorial plane and clearly part of the lobes.

On the far side of the red lobe, in addition to the polar hole, there exists a series of small indentations or irregularities. Such irregularities are not noticeable on the blue lobe. These irregularities may be the result of various instabilities in the thin Homunculus shell (see e.g. \citealt{smith13}) and/or interaction of the expanding Homunculus with surrounding pre-existing material \citepalias{steffen14}. Their exact cause is unknown at this time, but they do help to further distinguish the red and blue lobes, which is especially useful for tactile learners.

Finally, we note that the 3-D model of \citetalias{steffen14} lacks the ``dust skirt'' observed near the equatorial region in optical images. This is not a deficiency of the 3-D model, but rather a direct consequence of having modeled an H$_{2}$ infrared emission line. As noted in \citet{smith06}, there is no detectable near-infrared H$_{2}$ emission from the equatorial skirt. This is likely a result of both the low density of the skirt and its constant exposure to the hard UV flux from the secondary star \etaB, which spends most of its time during the 5.54-year orbit on the Earth-facing side of the nebula. The conditions within the skirt thus make the formation and survival of H$_{2}$ there extremely difficult, explaining its observed absence.

\subsection{The Colliding Stellar Winds and WWIR}\label{ssec:binaryres}
\vspace{2.0mm}
\subsubsection{Monochromatic 3-D Prints}

\begin{figure}
\centering \includemovie[
     3Dviews2=views_ApastronHighMdot.tex,
        toolbar, 
        label=Fig5.u3d,
     text={\includegraphics[width=84mm]{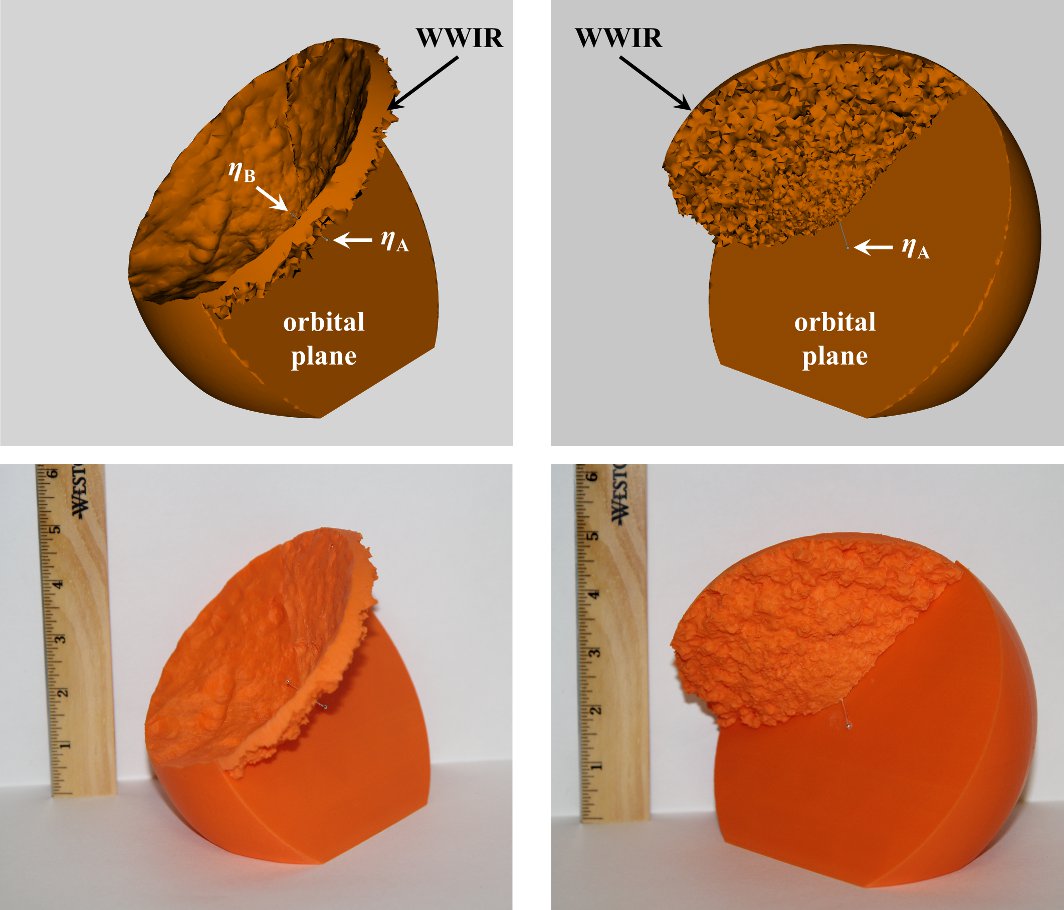}},
        3Dlights=CAD,
]{}{}{Fig5.u3d}
\caption{Comparison between the 3-D visualization (top row) and 3-D printed model (bottom row) for the SPH simulation at apastron. Columns show arbitrary views of the WWIR with the observer looking into the \etaB wind cavity (left), and with the \etaB cavity opening away from the observer (right). Both views look down on the orbital plane. Click the image for an interactive 3-D version of the model (Adobe Reader$^{\circledR}$ only). Pre-programmed views, available under the ``Views" menu in the 3-D model toolbar, include the projection of the system on the sky as viewed from Earth (view ``LOS", North up, East left), the primary wind and \etaB wind cavity only (view ``PrimaryWind"), and the WWIR plus stars only (view ``WWIR"). The diameter of the 3-D printed models is $\sim$6~inches (15.24~cm), as measured across the orbital plane through the center of the model. At this scale, 1~inch (25.4~mm) corresponds to $\approx 35.5$~AU $\approx 5.31 \times 10^{9}$~km. The locations of the stars, WWIR, and orbital plane are indicated in the 2-D images.}
\label{fig5}
\end{figure}

\begin{figure}
\centering \includemovie[
     3Dviews2=views_PeriastronHighMdot.tex,
        toolbar, 
        label=Fig6.u3d,
     text={\includegraphics[width=84mm]{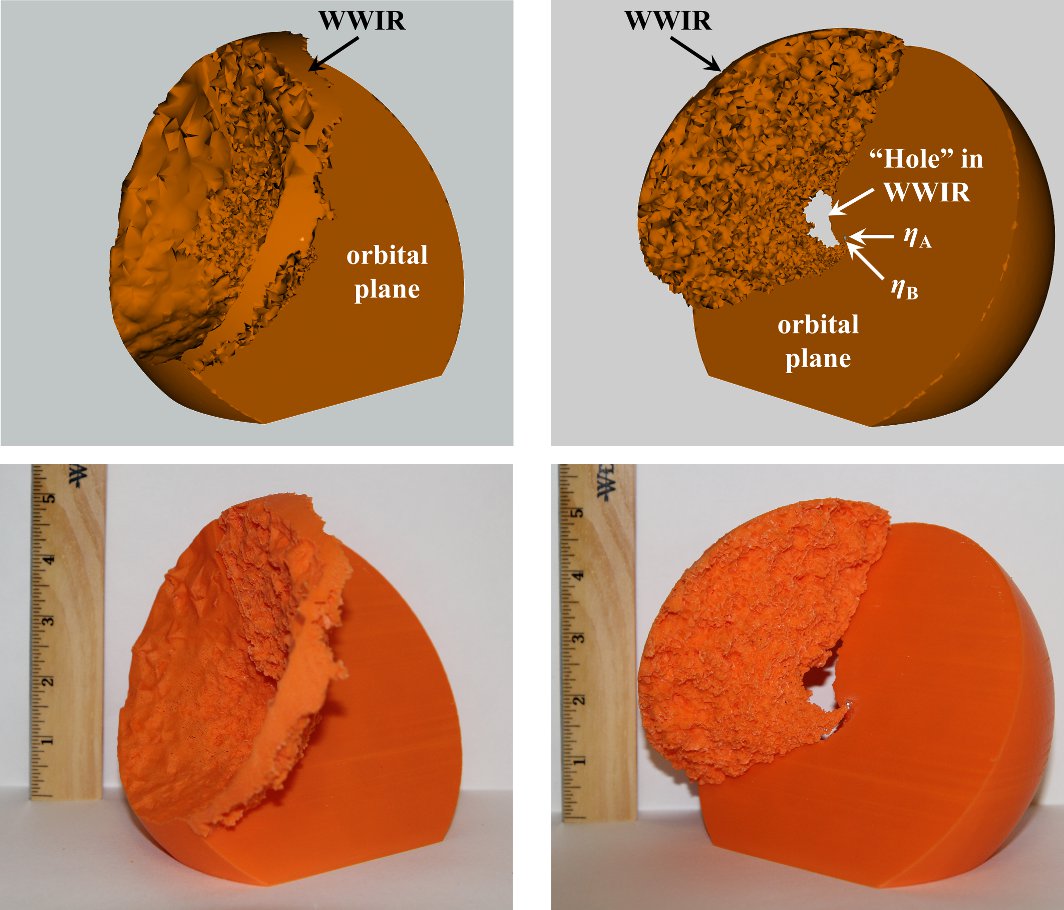}},
        3Dlights=CAD,
]{}{}{Fig6.u3d}
\caption{Same as Fig.~\ref{fig5}, but at periastron. Click the image for an interactive 3-D model (Adobe Reader$^{\circledR}$ only).}
\label{fig6}
\end{figure}

\begin{figure*}
\centering \includemovie[
     3Dviews2=views_Phase1p045HighMdot.tex,
        toolbar, 
        label=Fig7.u3d,
     text={\includegraphics[width=177mm]{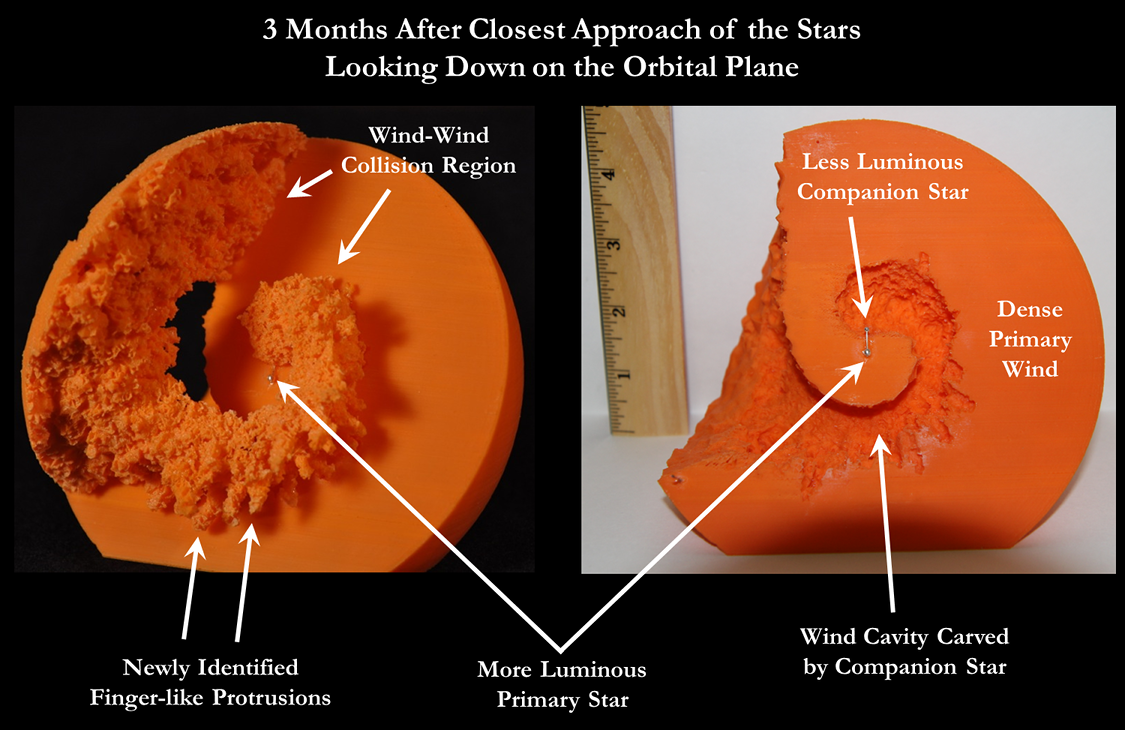}},
        3Dlights=CAD,
]{}{}{Fig7.u3d}
\caption{Detailed view of the two pieces that compose the monochromatic 3-D printed model at orbital phase 1.045 (3 months after periastron). The right panel shows the two pieces together, with the WWIR on top. The left panel shows the bottom half of the model only, with the stars, dense \etaA wind, and hollow \etaB wind cavity. The locations of the stars, \etaB wind cavity, primary wind, WWIR, and WWIR fingers are indicated. Both views are looking down on the orbital plane. Click the figure for an interactive 3-D model (Adobe Reader$^{\circledR}$ only).}
\label{fig7}
\end{figure*}

Figs.~\ref{fig5} -- \ref{fig7} present the monochromatic 3-D printing results. In each interactive 3-D figure, the default view has the model oriented at the same position on the sky as the \ec binary \citep{madura12}. In the ``Views'' menu of the 3-D graphics toolbar are options to display only the modified wind of \etaA, only the WWIR, or both together. The STL files and instructions on how to print these models are available on NASA's 3-D Resources website\footnote{\url{https://nasa3d.arc.nasa.gov/detail/eta-carinae-high}} and as online supplementary material to \citetalias{madura15}. Here we only briefly describe the key features of the 3-D prints. Details on the modeling and results can be found in \citetalias{madura15}.

Changes in the WWIR's shape and the distance of the WWIR from the stars are clearly conveyed in the 3-D prints. In some locations, the tetrahedral grid cells of the unstructured mesh can be seen. Even the small protruding fingers that extend radially from the spiral WWIR at orbital phase 1.045 are reproduced, although they are fairly delicate.

The ability to hold and inspect the 3-D models provides new perspective on the WWIR's geometry and an improved sense of the scale of the different dynamical structures. One appreciates more just how large the WWIR is compared to the stars and stellar separation. Moreover, the different textures on either side of the WWIR, caused by the differences in the density and temperature of the post-shock gas, and thus the grid cell size, provide tactile information about the differences between the adiabatic and radiatively-cooled sides of the WWIR. The WWIR is not a clean, smooth surface; rather, it is corrugated and contains small-scale protrusions that arise as a result of various instabilities. The 3-D prints provide tactile representations of these instabilities.

Apparent ``holes'' in the WWIR at periastron and three months after periastron indicate regions where there is actually no longer a wind-wind collision, a result of the highly elliptical orbital motion and decreased binary separation at these phases \citep{madura13,madura15}. Analysis of the $\delta$ parameter used to isolate the WWIR shows that in these holes, $\delta \lesssim 1$. The hole created near the WWIR apex at periastron also expands and propagates downstream along the WWIR's old trailing arm. Thus, the new spiral WWIR created just after periastron basically has no trailing arm, and consists mainly of a leading arm and structures above and below the orbital plane. 

Of course, perhaps the most interesting and noticeable of the 3-D printed structures are the WWIR fingers, which penetrate into the unshocked \etaA wind that is expanding on the periastron side of
the system. The fingers extend slightly above and below the orbital plane from the leading arm of the spiral WWIR, but interestingly they are not perfectly vertical (i.e. they are not perpendicular to the orbital plane). Instead, they point radially outward away from the stars and extend in the same direction that \etaB's fast wind is able to collide with \etaA's slower receding wind. The length of the fingers is also impressive; larger than the stellar separation at this phase ($\gtrsim$7~AU).

\subsubsection{Full-Color 3-D Print}

\begin{figure}
 \begin{center}
    \includegraphics[width=84mm]{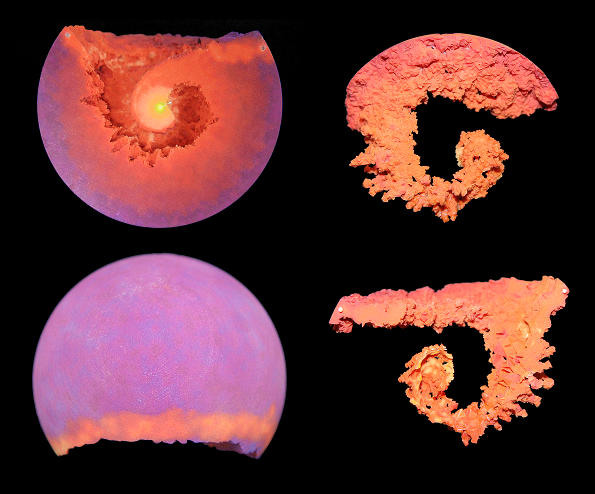}
    \caption{Full-color sandstone (FCS) 3-D print of \ec's binary colliding stellar winds and WWIR at orbital phase 1.045, based on the results of the 3-D SPH simulations of \citetalias{madura13}. Color shows logarithm of the density. The color bar is the same as that in Fig.~\ref{fig3}. \emph{Top left}: A view looking down on the orbital plane showing the dense \etaA winds and hollow \etaB wind cavity. \emph{Top right}: Top view of the spiral WWIR region, showing the finger-like protrusions. \emph{Lower left}: Bottom view of the model in the top left panel, showing the outer spherical edge of the pre- and post-shock \etaA winds (purple and pink, respectively). \emph{Lower right}: Bottom view of the WWIR showing the two pegs used to attach the WWIR to the \etaA winds portion of the model shown in the panels of the right column.}\label{fig8}
 \end{center}
\end{figure}

Fig.~\ref{fig8} presents pictures of our full-color 3-D print at orbital phase 1.045, with color showing logarithm of the density. The FCS printing process preserved remarkably well the colors as well as the model geometry. Comparison of the full-color 3-D print to the 3-D visualizations in Fig.~\ref{fig3} shows that the color table is, for the most part, faithfully reproduced. This includes the color gradient indicative of the $1/r^{2}$ drop off in density of the undisturbed portions of \etaA's wind. Amazingly, the change in color due to the density enhancement at the post-shock \etaA wind is preserved and visible around the boundary of the outer spherical edge of the bottom half of the model (bottom left panel of Fig.~\ref{fig8}). 

The FCS model is of a noticeably higher quality than the simple monochromatic MakerBot prints, which is expected due to the nature of the material (sandstone) and the professional audience of the printer, whose cost is approximately 30 -- 50$\times$ that of the MakerBot. The downside to the FCS print is it appears to be brittle and not quite as robust as ABS. One expects that if the FCS print were dropped, it would shatter on impact. FCS is not the ideal material for a 3-D print that is to be handled roughly or by lots of individuals, since delicate features could likely very easily break off (such as the WWIR fingers). Excessive handling and moisture can also potentially degrade the colors. Therefore, FCS would not be appropriate for use with visually impaired learners. This is not a major shortcoming though since full-color printing is likely not needed in such cases anyway.

\section{Discussion} \label{sec:disc}

\subsection{Relationship Between \ec's Homunculus Nebula and the Inner Colliding Stellar Winds}

The 3-D model of the Homunculus by \citetalias{steffen14} provides insights on the detailed structure of the bipolar lobes. The near-equatorial protrusions within each lobe and the spatially extended polar trenches are two significant discoveries by \citetalias{steffen14}. Of particular interest is the seemingly strong connection between the geometric properties of these features and the results of detailed numerical modeling of the central colliding winds \citep{madura13}. Both the trenches and protrusions are symmetric about the apastron direction of the binary orbit \citep{madura12}, with the angular distance between the protrusions ($\approx 110^{\circ}$) very similar to the angular extent of each polar trench ($\approx 130^{\circ}$, \citetalias{steffen14}). These angles are nearly identical to the total opening angle in the orbital plane of the current wind-wind collision cavity ($\approx 110^{\circ}$, using an \etaB/\etaA wind momentum ratio $\eta \approx 0.12$, \citealt{madura13}). 

The 3-D modeling further shows that the polar holes appear to be symmetrically offset from the polar axis of the Homunculus by $\sim$8$^{\circ}$ and aligned with the sky-projected apastron direction of the orbit \citepalias{steffen14}. The orbital axis of the binary is \emph{also} slightly offset from the Homunculus polar axis and nearly aligned with the sky-projected apastron direction \citep{madura12}. The polar holes may therefore be directly related to the binary orbital axis, especially since they and the orbital axis are closely aligned in 3-D space. Alternatively, the polar holes may be aligned with \etaA's rotation axis, if the stellar rotation and orbital axes are closely aligned.

These independent results, which turn out to have a strong correlation, hint toward a causal relation between the central binary colliding winds and the smaller geometric features of the Homunculus. It is noted that the 3-D reconstruction of the Homunculus was obtained without assuming the orientation of the orbit of the central binary \citepalias{steffen14}. Therefore, the interaction between the wind outflows and/or radiation from the central binary stars and their 3-D orientation in space has had, and possibly still has, a strong influence on the shape and structure of the surrounding Homunculus nebula \citepalias{steffen14}. 

The X-shooter observations and 3-D \Shape and SPH models thus pose serious complications for single-star models of the Homunculus's formation. Prior to the \Shape modeling of \citetalias{steffen14}, there was no strong reason to consider a binary model for the formation or shape of the Homunculus. Detailed 3-D hydrodynamical simulations of the Great Eruption and Homunculus's formation that include the central binary and its interacting winds are now required to determine if such interactions can indeed simultaneously reproduce the newly-identified small-scale features, overall bipolar shape, the polar holes, and the thin equatorial skirt.

\subsection{The Binary Colliding Winds and WWIR}

The features seen in 3-D prints of \ec's colliding stellar winds and WWIR may have some interesting implications for observational diagnostics of \ec and other highly-eccentric colliding wind binaries. The hole near the apex of the WWIR at periastron will affect the amount of shock-heated gas responsible for \ec's observed time-variable X-ray emission \citep{corcoran10,hamaguchi07,hamaguchi14}. The 3-D models imply that there should be no thermal X-ray emission from along the trailing arm at periastron. The WWIR hole also affects the amount and type of material in line-of-sight to the stars at periastron. At periastron, the column density of material between us and the stars will be dominated by unshocked \etaA wind material that is flowing to fill the wind cavity carved by \etaB during the broad part of the orbit. This makes `wind-eclipses' by \etaA of various observed features (e.g. X-rays, \citealt{corcoran10} and \HeII emission, \citealt{teodoro16}) somewhat easier to achieve at periastron, since the lack of a WWIR trailing arm allows the dense \etaA wind to expand and fill the \etaB wind cavity in line-of-sight, increasing the size of the eclipsing \etaA wind photosphere.

Observational implications of the WWIR fingers that form after periastron are unclear at this time. Unfortunately, they are too small to spatially resolve, even with \emph{HST}. Hot gas located within the fingers may produce X-rays, but the intensity and detectability of any such X-ray emission is unknown. It is also unclear if the fingers could contribute to the \HeII $\lambda$4686 emission observed across \ec's periastron passage \citep{teodoro16}. Portions of the fingers located within \etaA's inner He$^{+}$ zone may be able to produce a small amount of \HeII emission if the required He$^{+}$ is present. However, this is very speculative without more detailed modeling.

\citetalias{madura15} suggest that the fingers arise as a result of instabilities at the interface between the two colliding wind shocks, which undergo rapid complex changes around periastron due to the high orbital eccentricity and changing wind directions. The situation is analogous to that which occurs when a fast stellar wind interacts with a surrounding slower-moving circumstellar shell \citep[e.g.][]{toala11, vanmarle12}, with the post-shock \etaA wind forming a thin, dense shell via radiative cooling, and the post-shock \etaB wind remaining hot and cooling adiabatically (for details, see \citetalias{madura15}).

\subsection{Using the 3-D Prints for Outreach and Education}

We find that the 3-D print models are extremely useful as a visual aid to help explain to non-\ec experts, and even non-astronomers, the 3-D geometry and dynamics of the \ec Homunculus nebula, central binary, and WWIR. The 3-D prints are useful for illustrating concepts, relationships, and properties that are not easily conveyed by 2-D, or even 3-D, graphics. This is especially true when communicating with primarily tactile learners like the visually impaired.

The results of this work have garnered a significant amount of interest from the popular press and public, including two NASA press releases, a press conference at the 225th meeting of the American Astronomical Society, and the NASA Astronomy Picture of the Day on July 17, 2014. The works of \citetalias{steffen14} and \citetalias{madura15} have been featured in well over 100 on-line and print articles, including articles by Nature News, Science News, CNET.com, National Geographic News, BBC News, NBC News, Space.com, MIT Technology Review, 3Dprint.com, NASA Cutting Edge, Wired, and Sky \& Telescope. The NASA Goddard Space Flight Center (GSFC) also tracks total YouTube and NASA Science Visualization Studio (SVS) views to associated press-release web content. The narrated video created for the NASA release ``NASA Observatories Take an Unprecedented Look into Superstar Eta Carinae\footnote{\url{http://www.nasa.gov/content/goddard/nasa-observatories-take-an-unprecedented-look-into-superstar-eta-carinae/}}'', associated with the reveal of the \ec colliding wind 3-D prints, exceeded 140,000 views on the Goddard YouTube channel during its first week of release. The NASA SVS page housing videos and images of the 3-D prints received an additional 60,000 views during its first month.

In total, the YouTube video for the NASA press release ``Scientists Create First Full 3D Model of Eta Carinae Nebula,'' associated with the \citetalias{steffen14} 3-D model of the Homunculus, has received 59,578 lifetime views, while the ``NASA Observatories Take an Unprecedented Look into Superstar Eta Carinae'' YouTube video has received 257,359 lifetime views to date. NASAViz iPad stories and SVS pages associated with these press releases have received an \emph{additional} 545,094 lifetime views. We note that these numbers should be regarded as \emph{minimums} since other press organizations will host the NASA videos on their own websites and NASA does not collect statistics on the number of views received by these individual news outlets. As a result of this media attention, we have discovered that great interest also exists in using the 3-D prints to help encourage young students to pursue astronomy. This includes direct email messages from parents requesting copies of the 3-D prints for their children who are actively pursuing astronomy (the youngest of which was five years old). 

In order to make the 3-D print models more accessible to the public, we have made the monochromatic \ec 3-D models and instructions on how to print them freely available on NASA's 3D Resources website. Numerous individuals have successfully downloaded and 3-D printed the models, as evidenced by the availability of the Homunculus on the 3-D printing marketplace Shapeways\footnote{\url{http://www.shapeways.com/}} and a YouTube video in which a professor at the University of Nottingham 3-D printed a Homunculus out of titanium\footnote{\url{https://www.youtube.com/watch?v=IAZkpyFEbLg}}.   

Another obvious use for astronomical 3-D printing is outreach and education for the visually impaired. Our first attempt at using the \ec 3-D models for astronomy education with visually impaired students occurred during the summer of 2015. A complete set of three colliding wind models (one at each relevant phase) and several Homunculi were sent to the New Mexico Museum of Space History for use in a joint program with the New Mexico School for the Blind and Visually Impaired, who hosted a week long camp for visually impaired students from across the state to learn important life skills and astronomy. Feedback from students and educators who used the models was overall very positive\footnote{See e.g. \url{http://www.alamogordonews.com/story/news/local/2015/07/17/museum-adds-models-visually-impaired/32447733/}}.

\begin{figure}
 \begin{center}
    \includegraphics[width=82mm]{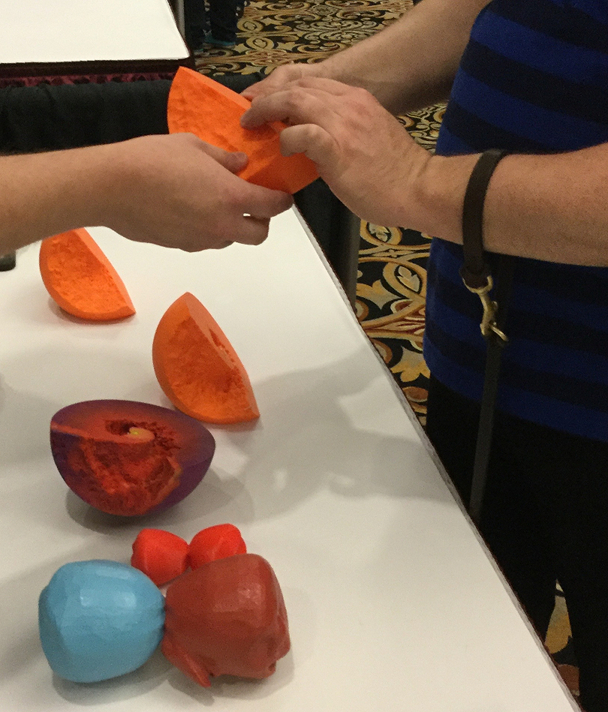}
    \caption{Picture of a visually-impaired attendee (right) to the 2016 Vision Rehabilitation and Assistive Technology Expo (VRATE), held in Glendale, AZ, on September 30, 2016, examining one of the 3-D prints of \ec's colliding stellar winds that was on display in the conference's exhibition hall.  Also visible in the image are an FCS print of the colliding winds (see e.g. Fig.~\ref{fig8}) and a red-blue Homunculus 3-D print (see Figs.~\ref{fig2} and \ref{fig4}).}\label{fig9}
 \end{center}
\end{figure}

Our second use of the \ec 3-D print models was at the 2016 Vision Rehabilitation and Assistive Technology Expo (VRATE), held in Glendale, AZ, on September 30, 2016. The Expo provided an excellent opportunity to present our work to approximately 200 -- 300 visually-impaired audience members. Following an hour-long morning keynote presentation explaining the background of the project, \ec, and the 3-D prints, a series of \ec Homunculi and 3-D colliding winds models were placed on display at a manned exhibit booth for approximately five hours. During this time, roughly two dozen visually-impaired expo attendees with a wide variety of backgrounds and ages interacted with the models, asking questions and providing helpful feedback (Fig.~\ref{fig9}).

The above and other interactions have provided us with valuable information and feedback crucial for improving these and other astronomical 3-D models targeted for use by the visually impaired. First, we discovered that it is generally necessary to have a subject matter expert present to describe the 3-D prints to the user. We found that the most helpful tactic was to guide the fingers of the user along the model while simultaneously explaining the key features. Such interactions typically took about 10 -- 15 minutes, sometimes up to 45 minutes, depending on the complexity and number of prints being explored, the number of questions asked, and the level of interest of the user. This leads to the obvious difficulty that a subject matter expert will usually not be available to the user, thus, an alternative should be supplied to maximize the educational impact of the models. One alternative is to include unique intuitive textures along the surface of the 3-D print that signify distinct features (e.g. lines for gas, stipple pattern for dust, spheres for stars, etc.), along with a separate ``texture key'' in braille. An alternative or additional method is to provide a pre-recorded audio file with each 3-D print that explains to the user the features being felt and their meaning. Ideally, one would like some way for the audio and 3-D print model to be linked, so that audio plays as the user runs their fingers across specific features. Advanced 3-D print models that include tactile sensors or buttons and embedded audio are one possible solution in development.

We wish to point out that audio should not be the sole method used for explaining the 3-D prints to visually impaired users, an important fact learned at VRATE when we encountered attendees that were both blind and deaf who wanted to experience the 3-D print models. We were fortunate in our case to have experienced volunteers present who could relay what we were explaining to these users via tactile sign language, tactile fingerspelling, or specific tactile technology such as a screen braille communicator or braille notetaker. A tactile braille explanation should thus ideally be developed to accompany any 3-D model developed.

We have also discovered that, depending on the quality of the 3-D printer, some amount of post processing of the 3-D print may be required to help visually impaired users avoid misconceptions when interacting with the print. This is mainly an issue with desktop FDM (Fused Deposition Modeling) 3-D printers that print successively layer-by-layer. This layer-by-layer process can often lead to a poor finishing quality and surface ridges that may be misinterpreted by visually-impaired users as actual physical features in the model. Unfortunately, such surface ridges can also lead to possible dimensional inaccuracies in the print model.  Both of these problems, luckily, can be resolved with various common 3-D printing post processing methods (sanding, tumbling, polishing, etc.).

Many visually-impaired users stated that the more small details provided on a 3-D print, the better their mental visualization and understanding. Several users surprisingly preferred the \ec colliding wind models over the Homunculus model. We were originally concerned that the colliding wind models would be too complex and difficult to understand, and that users would prefer the simpler and smoother Homunculus model. However, we were informed that overall, the Homunculus model was too smooth, and that extra detail, perhaps a textured surface, might help. Fortunately, users were still able to identify the key global features of the Homunculus model, including its overall bipolar shape, the near-equatorial protrusions, and the polar holes and trenches. Additionally, some users suggested adding a surface texture to the flat, smooth surface that defines the orbital plane in the colliding wind models. We therefore plan to add to future models a surface texture in the orbital plane that represents the $1/r^{2}$ drop off in density of the primary star's wind.

Related to the above, we also received feedback regarding the overall size of the models. Users commented that the colliding wind prints, at $\sim 6$~inches in diameter, were of adequate size to distinguish key details. Users preferred a larger ($\sim$7.5~inches from pole-to-pole) version of the Homunculus model as it more easily allowed them to feel and identify features. Overall, it was clear that users favored larger models that allow them to feel intricate details. The ability to distinguish small details can be critical in helping users understand complex situations. The physical size of the overall model and the ability to detect important small details should be considered key components of the design of any 3-D print model. Therefore, larger prints may need to be created in pieces and assembled whenever the printer volume is too small to produce the desired model size.

Our outreach and education work with the visually impaired continues with the development of a joint program with the South Carolina Commission for the Blind, scheduled to take place in June of 2017. As part of a multi-week program, a group of professional astronomers, engineers, educators of the visually impaired, and visually impaired scientists/engineers will work together to teach astronomy to visually impaired high school students. The details of this program are still in development and beyond the scope of this paper, but three curricular activities based on the \ec 3-D prints are planned now that we have evidence that our 3-D models are usable and understandable. These activities include (1) a 30 -- 45 minute Homunculus Nebula activity that consists of identifying structures in the bipolar Homunculus and covers topics such as size/scale in the Universe, physical origins of nebulae, stellar evolution, and binary stars; (2) a 1 -- 1.5 hour Colliding Stellar Winds activity that covers topics such as eccentric orbits, massive stars, stellar mass loss, and stellar evolution; and (3) a 20 -- 30 minute \emph{HST} Spectra activity that focuses on using stellar spectra of \ec to learn about stellar temperature, radiation, and spectra. Data gathered from the activities will be used to improve the design of future 3-D models for visually impaired learners, and to assess the effectiveness of the 3-D printed models to teach specific astronomy concepts.

\subsection{Current Limitations}

No developing technology is without limitations, and 3-D printing is no exception. Here we discuss some of the current key limitations of using 3-D printing for visualization. Most of these are fundamental to 3-D printing in general, but some could prove to be especially problematic depending on the specific application or intended use of the technology.

The first issue to mention is that of equal access and cost. While consumer-grade 3-D printers have decreased in price substantially, 3-D printing may still be too expensive for some (e.g. inner-city schools, developing nations, etc.). One possible remedy is the use of an on-line 3-D printing service such as Shapeways, but again, there is the associated cost of using such a service, plus the fact that such services, while widely available in the USA and Europe, are not yet available everywhere. A sustainable solution for providing near-equal access to astronomical 3-D print models needs to be developed. Hopefully with time, the cost of 3-D printing will drop to a level where everyone can benefit from the technology.

Similarly, if one wishes to use more advanced 3-D printing technologies such as full-color or transparent materials printing, a professional 3-D printer must be used. Again, an on-line 3-D printing service is one solution to this problem, since professional 3-D printers can cost hundreds of thousands of dollars. For now, there is no ideal solution to this problem, as one otherwise needs either a significant funding source to acquire a professional-grade 3-D printer, or one must wait for advanced 3-D printing features to become more affordable.

A second limitation of current 3-D printers is that of model size, both at the small and large ends. On the small end, one must be aware that all 3-D printers have a minimum build size, resolution, and accuracy. This limits the smallest possible feature size on any model. Thus, an extremely-high-resolution data set, simulation, or feature may not be accurately reproduced by a 3-D printer. One possible solution to this problem is to scale-up any key high-resolutions features and 3-D print them as a larger, separate model. Alternatively, one could increase the total size of the model itself. Therefore, unlike a 3-D visualization, it may be difficult or impossible to ``zoom'' in or out on a 3-D print model.

Similarly, the maximum size of a 3-D print is limited by the available build volume of the 3-D printer. Large models must be printed in separate pieces and then assembled. This introduces a new series of design challenges as extra care must be taken during the mesh creation phase to ensure that all components fit together properly after printing. Assembling multiple pieces also unfortunately introduces ``seems'' between the different components of the assembled model that can detract from the appearance and feel of the model, possibly introducing errors or misconceptions, especially for visually impaired users. This can be remedied with some degree of post-processing (adding filler, sanding, polishing, etc.), which is not necessarily complicated or tedious, but it may still be considered a disadvantage of 3-D printing. There is the added complication that if a low-grade desktop 3-D printer is used, the separate model pieces may not fit together properly at all due to insufficient print accuracy.

One major benefit of 3-D printing is that it provides maximum customization potential. However, this leads to another limitation, the fact that typically only one model can be printed at a time. 3-D printing thus lacks an economy of scale. Unlike conventional manufacturing, where mass production often leads to a lower per-unit cost, with 3-D printing, because every model is printed individually, there is no significant cost benefit when 3-D printing large numbers of objects. This makes 3-D printing large quantities of a model expensive, both in terms of cost and time (unless one is fortunate enough to have access to multiple 3-D printers). One solution is to create a mold of a specific final model that can be used for mass manufacturing by more conventional means. This way, many models can be made for dissemination to a large audience once a unique 3-D model has been prototyped using a 3-D printer.

One other possible disadvantage when using a consumer-grade desktop 3-D printer is the inability to show multiple surfaces simultaneously. 3-D visualizations permit volume renderings, transparency, and cross-sectional slicing, and the ability to move rather quickly from one to another. With a desktop 3-D printer, one is usually limited to individual solid surfaces, and multi-component models consisting of multiple isosurfaces and/or colors again requires multiple parts and assembly. However, 3-D prints in general \emph{are} capable of showing more than one surface at a time. 3-D printing using transparent materials is now possible \footnote{E.g. \url{http://blog.luxexcel.com/transparent-3d-printing/transparent-3d-print/}}, so that “nested” 3-D prints can be created that show different surfaces or components\footnote{See e.g. the 3-D printed transparent model of the human liver at \url{https://www.youtube.com/watch?v=Gjzp5ehjMA0}}.  Most of these advanced technologies have yet to be applied to astronomical models, but they are prevalent in e.g. medicine \citep{ventola14}.  The real issue for such complicated prints is cost, since only more expensive printers are capable of using transparent materials at this time.  Transparency and color are irrelevant for models intended for use by the visually impaired, so instead, different surface textures are usually required if multiple similar shapes are to be used in one complex multi-part model.

\section{Summary \& Conclusions} \label{sec:summ}

We have demonstrated that 3-D printing is a powerful tool that can be used to visualize a range of multi-D astronomical datasets, from observations to numerical simulations. There is a rapidly growing necessity to move beyond the limits of simple 2-D displays, especially as the number of multi-D datasets from IFS instruments and computational simulations continues to grow. Uncovering new ways to visualize, understand, and communicate the contents of such multi-D datasets is fundamental if we are to make sense of our intrinsically multi-D Universe. 3-D printing could also prove to be vital to how astronomer's reach out and share their work with each other, the public, and new audiences.

3-D printing moves well beyond interactive 3-D graphics and provides an excellent tool for both visual and tactile learners, since 3-D printing can now easily communicate both complex geometries and full color information. Some of the limitations of interactive 3-D graphics are also alleviated by 3-D printable models, including issues of limited software support, portability, accessibility, and sustainability. Some current limitations of 3-D printing include limits on model size, a lack of an economy of scale, and the difficulty and cost associated with producing complicated nested, full-color, or transparent models.

We have shown that there is public interest in astronomical 3-D printing, and that 3-D printing can be used to guide young students' budding interest in astronomy, as well as help the visually impaired build mental images of the Universe around them. 3-D printing has helped us share the results of various 3-D modeling efforts of \ec, including the detailed structure of the Homunculus nebula and the central binary colliding stellar winds. In addition to helping visualize distinct features, the 3-D prints can be used to illustrate the possible impact of the binary interacting winds on the shape of the surrounding Homunculus nebula. We also use \ec as an example to demonstrate how the addition of color can improve the usefulness of 3-D printing, conveying another layer of information that can help guide sighted users.

Attached to this article are full-color 3-D printable files of both the red-blue Homunculus model and the colliding stellar winds/WWIR at orbital phase 1.045. Any readers with questions about 3-D printing the models, the model design, and/or the file creation process are welcome to contact us directly. We also welcome anyone interested in using 3-D printing to adapt their research for outreach and education with the visually (and/or non-visually) impaired to contact us.\\

\acknowledgments
\section*{Acknowledgments}
Support for Program number HST-AR-14301 was provided by NASA through a grant from the Space Telescope Science Institute, which is operated by the Association of Universities for Research in Astronomy, Incorporated, under NASA contract NAS5-26555. I thank the referee, Wolfgang Steffen, for his very helpful suggestions.


\end{document}